\documentclass{emulateapj}

\usepackage{natbib}
\shorttitle{Metallicity and Environment in the SDSS}
\shortauthors{Cooper et al.}

\begin{document}


%\title{The Relationship between Environment and Metallicity: Environment's
%  Role in the Mass--Metallicity Relation}
\title{The Role of Environment in the Mass--Metallicity Relation}

%The Relationship between Star
%  Formation and Galaxy Environment at $\lowercase{z} < 1$}


\author{
Michael C.\ Cooper\altaffilmark{1,2},
Christy A.\ Tremonti\altaffilmark{1,3},
Jeffrey A.\ Newman\altaffilmark{4},
Ann I.\ Zabludoff\altaffilmark{1}
}

\altaffiltext{1}{Steward Observatory, University of Arizona, 
933 N.\ Cherry Avenue, Tucson, AZ 85721 USA; 
cooper@as.arizona.edu, tremonti@as.arizona.edu, azabludoff@as.arizona.edu}

\altaffiltext{2}{Spitzer Fellow}

\altaffiltext{3}{Hubble Fellow}

\altaffiltext{4}{Department of Physics and Astronomy, University of
  Pittsburgh, 401--C Allen Hall, 3941 O'Hara Street, Pittsburgh, PA 15260
  USA; janewman@pitt.edu}

\begin{abstract}

  Using a sample of 57,377 star--forming galaxies drawn from the Sloan
  Digital Sky Survey, we study the relationship between gas--phase oxygen
  abundance and environment in the local Universe. We find that there is a
  strong relationship between metallicity and environment such that more
  metal--rich galaxies favor regions of higher overdensity. Furthermore,
  this metallicity--density relation is comparable in strength to the
  color--density relation along the blue cloud. After removing the mean
  dependence of environment on color and luminosity, we find a significant
  residual trend between metallicity and environment that is largely driven
  by galaxies in high--density regions, such as groups and clusters. We
  discuss the potential source of this relationship between metallicity and
  local galaxy density in the context of feedback models, with special
  attention paid to quantifying the impact of environment on the scatter in
  the mass--metallicity relation. We find that environment is a
  non--negligible source of scatter in this fundamental relation, with
  $\gtrsim \! 15\%$ of the measured scatter correlated with environment.

\end{abstract}

\keywords{galaxies:evolution, galaxies:statistics, galaxies:
  abundances, galaxies:fundamental parameters, large--scale structure of
  universe}

\section{Introduction}
\label{sec_intro}

Gas--phase metallicity is one of the most fundamental characteristics of a
galaxy, affecting the evolution of its stellar population and the
composition of its interstellar medium (ISM). Moreover, metallicity
indirectly traces a galaxy's star--formation history and reflects the
balance of several important physical processes: the release of metals into
the interstellar medium via supernovae and stellar winds, the ejection of
gas via galactic outflows, and the accretion of gas onto the galaxy from
the surrounding environs. Understanding how metallicity evolves, especially
in relation to other fundamental galaxy properties, is essential in
isolating the physical mechanisms that drive star formation and, more
generally, galaxy evolution.

As first observed by \citet{lequeux79}, metallicity is strongly correlated
with galaxy stellar mass, such that more massive galaxies are more
metal--rich in composition. Due to the relative ease of measuring
luminosities versus stellar masses, many subsequent studies extended this
early work to larger samples of galaxies by studying the correlation
between luminosity and metallicity \citep[e.g.,][]{skillman89, brodie91,
  zaritsky94, garnett02, kobulnicky03, lamareille04}. Using large data sets
from surveys such as the Sloan Digital Sky Survey \cite[SDSS,][]{york00},
more recent analyses have brought measurements of the luminosity-- and
mass--metallicity relations on par with each other, measuring relations
that span more than ten magnitudes in optical luminosity and six orders
of magnitude in stellar mass, ranging from dwarf galaxies up to the most
massive star--forming systems \citep[e.g.,][]{pilyugin00, lee03,
  tremonti04, shapley05, erb06, lee06}.

Both the luminosity--metallicity and mass--metallicity relations show
significant scatter, with only half the observed spread in the metallicity
distribution at fixed stellar mass being due to observational error and an
even greater ($\sim \! 50\%$ greater) scatter measured for the
luminosity--metallicity relation \citep{tremonti04}.  Various studies have
pointed to physical sources of the scatter in these fundamental
relations. For example, studying a sample of UV--selected galaxies at $z <
0.4$, \citet{contini02} find that these systems are offset from the
luminosity--metallicity relation due to a recent starburst that has
enriched their ISM and decreased their mass--to--light ratios, moving them
off of the median trend. As illustrated by \citet{tremonti04}, however,
these results suggest that the relationship between metallicity and stellar
mass (and not luminosity) is more fundamental; even when accounting for
variations in mass--to--light ratio due to dust attenuation and observing
at redder wavelengths so as to minimize the impact of newly formed stars on
the measured luminosity, the scatter in the luminosity--metallicity
relation is still greater than that observed between stellar mass and
metallicity.

By analyzing the correlations between the scatter in the mass--metallicity
relation and other galaxy properties (e.g., rest--frame color, inclination,
photometric concentration, etc.), \citet{tremonti04} point to a potential
connection with stellar surface mass density, $\mu_*$, such that galaxies
with higher surface densities are more metal--rich relative to galaxies of
similar stellar mass \citep[see also][]{ellison08a}. This trend is
potentially explained by a scenario where galaxies with higher surface
densities have converted more of their gas reservoirs into stars and
thereby elevated their metallicity. In conflict with this picture, however,
\citet{tremonti04} find no significant correlation between scatter in the
mass--metallicity relation and morphology (as traced by the concentration).

Local galaxy density (i.e., the local ``environment'') could act as an
alternate source of the scatter in the mass--metallicity
relation. Supernovae are predicted to enrich the intergalactic medium (IGM)
over roughly Mpc scales \citep[e.g.,][]{adelberger05}, which would impact
the metallicity of nearby galaxies in high--density
environments. Similarly, in clusters of galaxies, intracluster supernovae
may inject a significant quantity of metals into the intracluster gas
\citep{domainko04}, which is subsequently accreted onto cluster members,
thereby raising their metallicity.

Galaxies in high--density regions should collapse and form stars earlier
than their counterparts in low--density environs; studies of the
color--density relation show that galaxies with older stellar populations
favor higher--density environments at $z \sim 0$ \citep[e.g.,][]{
  balogh04a, balogh04b, blanton05a} and at $z \sim 1$
\citep[e.g.,][]{cooper06, smith05}. Thus, galaxies might be expected to
become more metal--rich sooner in high--density regions.

Direct evidence for the potential role of environment in shaping the
metallicity of a galaxy is found in observational work by \citet{kewley06},
which shows that galaxy interactions, common in galaxy pairs and groups
\citep{cavaliere92}, may lead to inflows that drag metal--poor gas to the
galaxy center, decreasing the gas--phase metallicity in such systems
\citep[see also][]{ellison08b}. The analysis of \citet{kewley06}, however,
probes a limited range of extreme environments (focusing on pairs at close
separations), which provides a vastly incomplete view of the role of
environment. Similarly, analysis of 41 metal--rich, low--mass galaxies by
\citet{peeples08}, finds that such outliers on the mass--metallicity
relation tend to be isolated and undisturbed systems (i.e., reside in
low--density environments). Though, this work is clearly limited by its
small sample size and the restricted mass range probed.

In this paper, we utilize data from the Sloan Digital Sky Survey to study
the relationship between metallicity and environment among the nearby,
star--forming galaxy population. Specifically, we inspect the correlation
between metallicity and environment in comparison to well--established
correlations between environment and properties such as rest--frame
color. In addition, we examine the potential impact of environment on the
scatter in the mass--metallicity relation. In \S \ref{sec_data}, we outline
the data set used in this analysis. In \S \ref{sec_res1}, \S \ref{sec_rem},
and \S \ref{sec_res2}, we present our results on the relationship between
metallicity and environment at $z \sim 0.1$. We then endeavor to quantify
the role of environment in driving the scatter in the mass--metallicity
relation in \S \ref{sec_scatter}. In \S \ref{sec_disc} and \S
\ref{sec_sum}, the results of this analysis are then discussed and
summarized. Unless otherwise noted, all work in this paper employs a flat,
$\Omega_{\Lambda} = 0.7$, $\Omega_{m} = 0.3$, $h=1$ cosmology.

\section{Data Sample}
\label{sec_data}

To study the relationship between local galaxy environment and various
galaxy properties, including metallicity, we utilize data drawn from the
SDSS public data release 4 \citep[DR4,][]{adelman06}, as contained in the
NYU Value--Added Galaxy Catalog \citep[NYU--VAGC,][]{blanton05b}. We
restrict our analysis to the redshift regime $0.05 < z < 0.15$ in an effort
to probe a broad range in galaxy luminosity, with large sample size, while
minimizing aperture effects related to the finite size (3'') of the SDSS
fibers. In addition, we limit our sample to SDSS fiber plates for which the
redshift success rate for targets in the main spectroscopic survey is 80\%
or greater.

\subsection{Measurements of Local Galaxy Environments}
\label{sec_data_environ}

We estimate the local galaxy overdensity, or ``environment'', in the SDSS
using measurements of the projected $3^{\rm rd}$--nearest--neighbor surface
density $(\Sigma_3)$ about each galaxy, where the surface density depends
on the projected distance to the $3^{\rm rd}$--nearest neighbor, $D_{p,3}$,
as $\Sigma_3 = 3 / (\pi D_{p,3}^2)$. In computing $\Sigma_3$, a velocity
window of $\pm 1000\ {\rm km}/{\rm s}$ is employed to exclude foreground
and background galaxies along the line--of--sight. Tests by
\citet{cooper05} found this environment estimator to be a robust indicator
of local galaxy density within deep surveys.

To correct for the redshift dependence of the SDSS sampling rate, each
surface density value is divided by the median $\Sigma_3$ of galaxies at
that redshift within a window of $\Delta z = 0.02$; this converts the
$\Sigma_3$ values into measures of overdensity relative to the median
density (given by the notation $1 + \delta_3$ here) and effectively
accounts for redshift variations in the selection rate \citep{cooper05}.
Finally, to minimize the effects of edges and holes in the SDSS survey
geometry, we exclude all galaxies within $1\ h^{-1}$ Mpc (comoving) of a
survey boundary. For further details regarding the computation of galaxy
environments in the SDSS, we direct the reader to \citet{cooper06} and
\citet{cooper08}.

\subsection{Measurements of Rest--frame Color, Absolute Magnitude, and
  Stellar Mass}
\label{sec_data_photo}

We compute rest--frame $g-r$ colors, absolute $r$--band magnitudes $(M_r)$,
and stellar masses from the apparent, petrosian $ugriz$ magnitudes in the
SDSS DR4, using the {\it kcorrect} K--correction code (version v4\_1\_2) of
\citet[][see also \citealt{blanton03b}]{blanton07}. The template SEDs
employed by {\it kcorrect} are based on those of \citet{bc03}. To estimate
stellar masses, the best--fit SED given the observed $ugriz$ photometry and
spectroscopic redshift is used to directly compute the stellar
mass--to--light ratio $({\rm M}_{*}/L)$, assuming a \citet{charbrier03}
initial mass function. We have also employed the stellar mass estimates of
\citet{kauffmann03a}, which do not rely on fitting SEDs to the SDSS
photometry; instead they have been derived by fitting to stellar
absorption--line indices, measured from the observed SDSS spectra, while
also attempting to correct for attenuation due to dust. Using these
alternate stellar mass values produces no significant changes in the
results of our analyses. Finally, all magnitudes within this paper are
given in the AB system \citep{oke83}.
% Except for those in the nearly--AB SDSS system, all magnitudes in this
% paper are given in the AB system.

\begin{figure}[h!]
\centering
\plotone{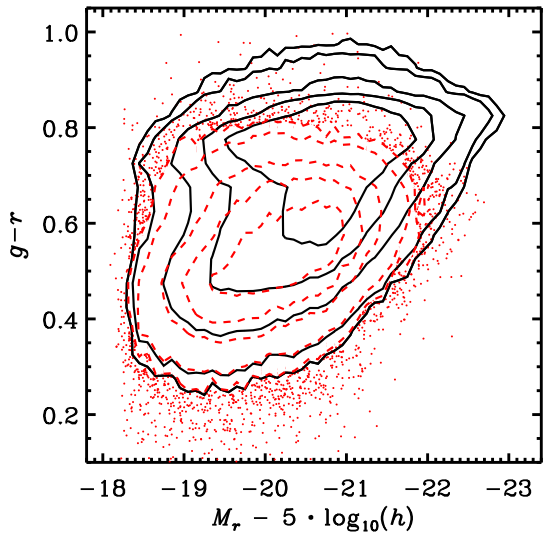}
\caption{The color--magnitude distribution for all 246,242 galaxies within
  $0.05 < z < 0.15$ in the SDSS NYU--VAGC DR4 catalog (\emph{black solid
    lines}) and for the 57,377 star--forming galaxies with accurate
  environment and metallicity measurements in the redshift range $0.05 < z
  < 0.15$ (\emph{red points and dashed lines}). The dominant impact of the
  cuts made in selecting our star--forming sample is the exclusion of
  quiescent galaxies and AGN, which preferentially reside at the red end of
  the blue cloud or on the red sequence.}
\label{cmd_fig}
\end{figure}

\subsection{Measurements of Spectral Properties: 
Metallicity and Star--Formation Rate}

\begin{figure}[h!]
\centering
\plotone{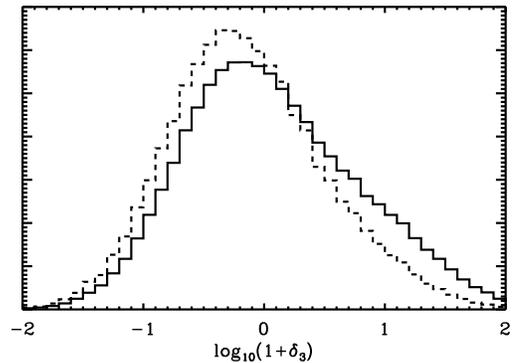}
\caption{The distribution of the logarithm of the local ovderdensities,
  $\log_{10}(1+\delta_3)$, for all 232,882 SDSS galaxies within $0.05 < z <
  0.15$ and with accurate environment measurements (\emph{solid line}) and
  for the 57,377 star--forming galaxies with accurate environment and
  metallicity measurements (\emph{dashed line}). Here, we scale the two
  histograms so that their integrals are equal. The star--forming galaxies
  are biased towards lower overdensities, though the sample still spans the
  full range of environments probed by the SDSS.}
\label{delta3_fig}
\end{figure}

To study the metallicities of the SDSS galaxies, we utilize oxygen
abundances, $12 + \log_{10}({\rm O}/{\rm H})$, from \citet{tremonti04},
which have been derived by statistically comparing the fits of nebular
emission lines in the SDSS spectra to the models of \citet{charlot01}. The
sample is limited to only those sources with H$\beta$, H$\alpha$, and [N
{\small II}] $\lambda6584$ all detected at a $5\sigma$ level. Furthermore,
we constrain our analysis to the star--forming galaxy sample, excluding
those objects hosting an active galactic nucleus (AGN) according to the
conservative line--diagnostic criteria of \citet{kauffmann03b}. By also
requiring accurate environment measures, as described above, we arrive at a
final star--forming galaxy sample including 57,377 sources at $0.05 < z <
0.15$. A distribution of the sample in color--magnitude space is shown in
Figure \ref{cmd_fig}. By excluding quiescent galaxies and active galactic
nuclei (AGN), the star--forming sample is biased against galaxies residing
on the red end of the blue cloud or the red sequence. In Figure
\ref{delta3_fig}, we also show the distribution of environment measures for
the star--forming sample relative to that for the full SDSS sample. While
the star--forming galaxies are biased towards lower overdensities
(consistent with a sample dominated by blue galaxies), the sample still
spans a full range of environments, from voids to clusters.

To probe the ongoing star--formation activity in this sample, we employ the
aperture--corrected star--formation rates (SFR) of \citet{brinchmann04},
which are estimated by fitting models to the nebular emission features in
the SDSS spectra. For the star--forming galaxy population, these SFRs show
excellent agreement with UV--based star--formation rate estimates
\citep{salim07}. Note that the SFR values of \citet{brinchmann04} are
estimated using $h=0.7$ rather than $h=1$.

\section{The Dependence of Mean Environment on Metallicity}
\label{sec_res1}

A wide variety of galaxy properties at low and intermediate redshift have
been shown to correlate with environment. For instance, at $z < 1$, blue,
star--forming galaxies are found to reside in regions of lower galaxy
density in comparison to red and dead systems \citep[e.g.,][]{balogh98,
  kauffmann04, cooper06, cucciati06, capak07}. Moving beyond direct studies
of the color--density or morphology--density relations, \citet{blanton05a}
analyzed the relationship between environment and the luminosities, surface
brightnesses, rest--frame colors, and structural characteristics
(S\'{e}rsic indices) of nearby galaxies in the SDSS sample. Among this set
of galaxy properties, they found that color and luminosity are the pair
that prove to be most predictive of the local environment. That is,
rest--frame color and luminosity are the two characteristics most closely
related to the galaxy density, as measured on small ($\sim 1\ h^{-1}$ Mpc)
scales. Furthermore, at fixed color and luminosity, they found no
significant trend between local galaxy density and surface brightness or
S\'{e}rsic index among the star--forming population --- although, for the
full SDSS sample, there is some residual correlation observed at high
luminosities, likely driven by rare, very luminous, red systems in dense
environments such as brightest cluster galaxies \citep{blanton05a}.

\begin{figure}
\plotone{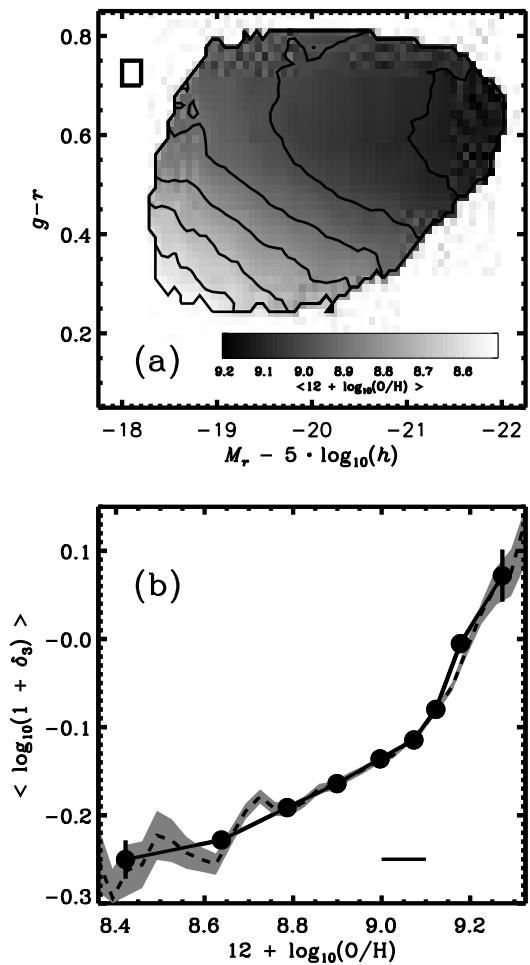}
\caption{(\emph{Top}) We plot the mean gas--phase metallicity, $12 +
  \log_{10}({\rm O}/{\rm H})$, as a function of rest--frame color and
  absolute magnitude, computed in a sliding box of width $\Delta M_r = 0.2$
  and height $\Delta (g-r) = 0.05$, as shown in the upper left corner. The
  mean metallicity depends on both color and luminosity, with more luminous
  and redder galaxies tending to have greater metal
  content. (\emph{Bottom}) We plot the mean galaxy overdensity as a
  function of gas--phase metallicity for the star--forming population. The
  dashed black line and grey shaded region show the mean and 1$\sigma$
  uncertainty in the mean overdensity computed in a sliding box of width
  $\Delta (12 + \log_{10}({\rm O}/{\rm H})) = 0.1$. The points and
  corresponding error bars give the mean and $1\sigma$ error in the mean in
  discrete bins of metallicity. We compute overdensities using the full
  SDSS galaxy sample (i.e., not just the star--forming population), thus
  the mean values plotted here are generally less than zero (in the
  logarithm), as expected from the color--density relation.}
\label{metal_fig}
\end{figure}

Like surface brightness and S\'{e}rsic index, metallicity is strongly
correlated with color and luminosity, such that brighter and redder sources
on the blue cloud tend to have higher metal abundances. This trend is
clearly evident in the top panel of Figure \ref{metal_fig}, where we show
the mean gas--phase oxygen abundance, $12 + \log_{10}({\rm O}/{\rm H})$, as
a function of rest--frame color and absolute magnitude for the SDSS
star--forming sample. Not surprisingly, when we examine the relationship
between metallicity and environment, we find a strong trend that includes
contributions from the correlations between (\emph{a}) metallicity, color,
and luminosity and (\emph{b}) color, luminosity, and environment. As shown
in the bottom portion of Figure \ref{metal_fig}, the typical environment
increases in overdensity for galaxies with higher
metallicities.\footnote{The local environment is thought to influence
  galaxy properties, such that galaxy properties are typically studied as a
  function of environment. In Figure \ref{metal_fig}b, however, we plot the
  dependence of mean environment on metallicity and not vice versa for one
  significant reason: measurements of environment are significantly more
  uncertain than measures of metallicity. Thus, binning galaxies according
  to local overdensity would yield significant correlation between
  neighboring environment bins, which would consequently smear out the
  underlying correlation between metallicity and local galaxy overdensity.}
This metallicity--environment relation agrees with the well--established
color--density relation along the blue cloud \citep{hogg03, blanton05a},
where the mean galaxy density increases with color.

\begin{figure*}[tb]
\plotone{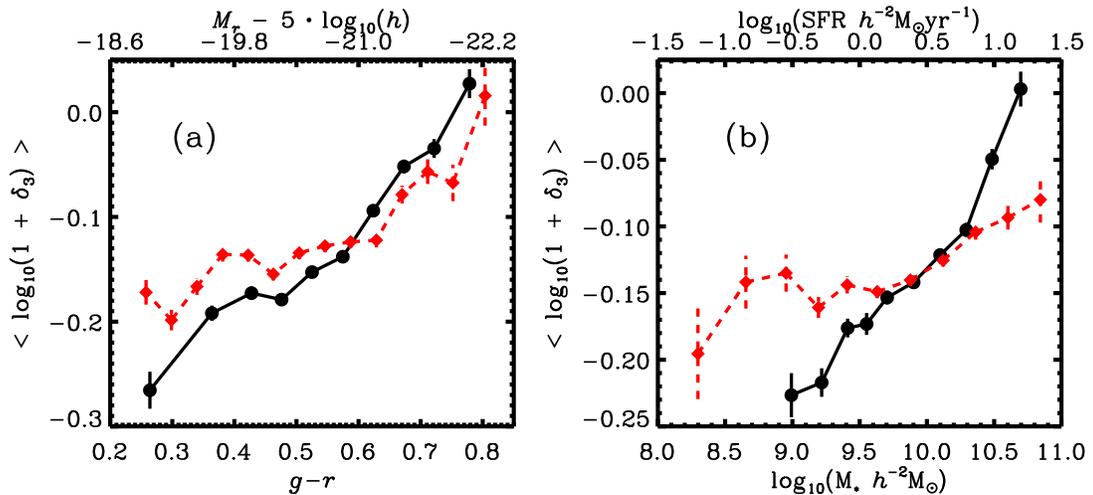}
\caption{(\emph{Left}) For the star--forming population, we plot the
  dependence of mean overdensity, $\log_{10}(1+\delta_3)$, on rest--frame
  color and absolute magnitude, as given by the black circles plus solid
  line and red diamonds plus dashed line, respectively. (\emph{Right})
  Similar to the plot on the left, but for stellar mass, ${\rm M}_{*}$,
  (black circles and solid line) and star--formation rate (red diamonds and
  dashed line). Within the star--forming sample, the
  metallicity--environment trend is as strong as the color--density
  relation, which is the strongest of the relations plotted here.}
\label{clss_fig}
\end{figure*}

While the trend evident in Figure \ref{metal_fig}b may not be surprising,
the \emph{strength} of this environment--metallicity relation is very
striking when compared to that seen between environment and color,
luminosity, stellar mass, or star--formation rate. As shown in Figure
\ref{metal_fig}b and Figure \ref{clss_fig}, the metallicity--density
relation is roughly comparable in strength to the color--density relation
amongst the star--forming population. In addition, the dependence of mean
overdensity on luminosity, stellar mass, and SFR are all weaker than that
observed with metallicity. Along the blue cloud, there is clearly a strong
relationship between gas--phase oxygen abundance and the local galaxy
environment.

\section{Removing the Mean Color--Luminosity--Environment Relation}
\label{sec_rem}

\begin{figure*}[tb]
\plotone{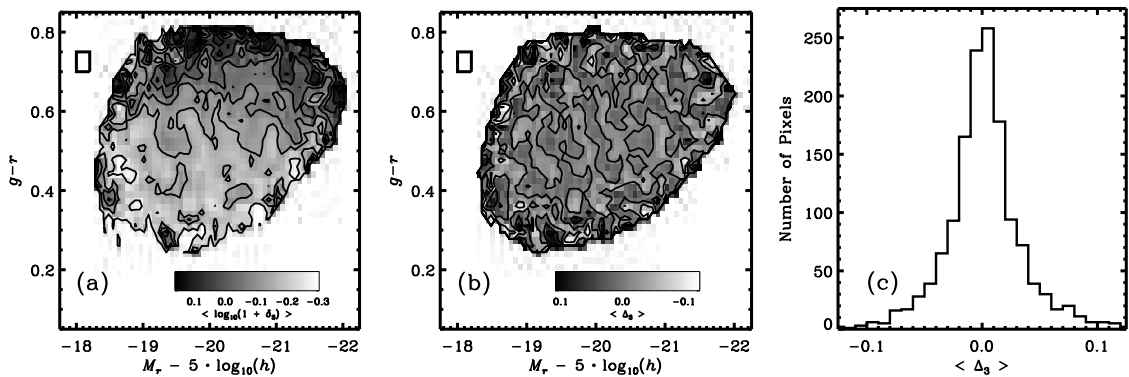}
\caption{(\emph{Left}) For the star--forming sample, we show the mean
  galaxy overdensity, $\log_{10}(1+\delta_3)$, as a function of rest--frame
  galaxy color, $g-r$, and absolute magnitude, $M_r$, computed in a sliding
  box of width $\Delta M_r = 0.2$ and height $\Delta (g-r) = 0.05$. The
  size and shape of the box are illustrated in the upper left corner of the
  plot. (\emph{Middle}) The mean residual environment, $\Delta_3$, as a
  function of color and magnitude, computed in the same sliding
  box. (\emph{Right}) We plot the distribution of mean residual environment
  for all regions where the sliding box contains 20 or more galaxies.)}
\label{mden_fig}
\end{figure*}

Given the relationships between metallicity, color, and luminosity, it
would be reasonable to expect that the strong relationship between local
galaxy density and metallicity is entirely contained in the
color--luminosity--environment relation (within the precision of our
measurements), such that there is no residual trend between metallicity and
environment at fixed color and luminosity --- similar to the findings of
\citet{blanton05a} for surface brightness and S\'{e}rsic index. To probe
the dependence of environment on metallicity at fixed color and luminosity,
we fit and remove (subtract) the dependence of mean environment on
rest--frame color and absolute magnitude. Figure \ref{mden_fig}a shows the
mean overdensity as a function of rest--frame $g-r$ color and $r$--band
absolute magnitude, or $< \log_{10}(1 + \delta_3)[g-r, M_r] >$, for the
SDSS star--forming sample. There is a clear color--density trend, where the
mean overdensity increases with color along the blue cloud. To remove this
relationship of environment to color and luminosity, we subtract the mean
overdensity at the color and luminosity of each galaxy from the measured
overdensity:
\begin{equation}
  \Delta_3 = \log_{10}(1 + \delta_3) - < \log_{10}(1 + \delta_3)[g-r, M_r]
  > ,
\label{D3_eqn}
\end{equation}
where the distribution of mean environment with color and absolute
magnitude, $ < \log_{10}(1 + \delta_3)[g-r, M_r] > $ (see Fig.\
\ref{mden_fig}a), is median smoothed on $\Delta(g-r) = 0.15$ and $\Delta
M_{r} = 0.6$ scales prior to subtraction.

An alternate method of effectively removing the
color--luminosity--environment relation from our analysis would be to study
the metallicity--environment relation in bins of rest--frame color and
absolute magnitude (or in bins of stellar mass). This approach, however,
can be far less sensitive, since dividing the sample into such restricted
subsets reduces the signal--to--noise ratio of any trend that occurs across
the entire color--magnitude distribution (i.e., spans the blue cloud). In
\S \ref{sec_disc_2}, we return to this point in comparison to other recent,
related analyses.

The ``residual'' environment, $\Delta_3$, quantifies the overdensity about
a galaxy relative to galaxies of similar color and luminosity, where values
of $\Delta_3$ greater than zero correspond to galaxies in environments more
overdense than the typical galaxy with like star--formation history (that
is, like $g-r$ and $M_r$). Figure \ref{mden_fig}b shows the dependence of
mean $\Delta_3$ on color and luminosity; no significant color or luminosity
dependence is evident. Furthermore, Figure \ref{mden_fig}c displays the
distribution of $< \! \Delta_3 \! >$ values from Fig.\ \ref{mden_fig}b,
illustrating that deviations from $< \! \Delta_3 \! > = 0$ are small.

\begin{figure}[h]
\plotone{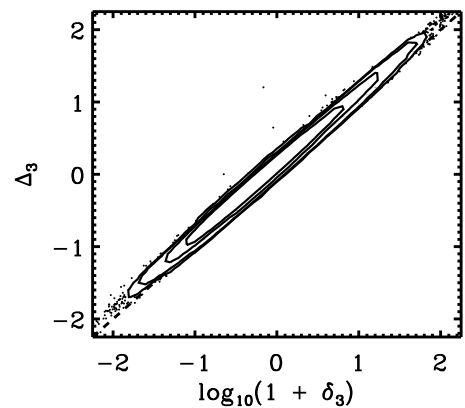}
\caption{For the 57,377 galaxies in the star--forming population, we plot
  the relationship between the ``residual'' environment, $\Delta_3$, and
  the ``absolute'' environment, $\log_{10}(1+\delta_3)$. For a definiton of
  the $\Delta_3$ statistic, refer to Equation \ref{D3_eqn}.}
\label{diff_fig}
\end{figure}

While the $\Delta_3$ statistic effectively removes the mean color--density
and luminosity--density relations from the data set, this measure of the
residual environment is only a small perturbation to the ``absolute''
overdensity, $\log_{10}(1+\delta_3)$. As shown in Figure \ref{diff_fig},
the $\Delta_3$ value for each galaxy in our sample is still strongly
correlated with the corresponding $\log_{10}(1+\delta_3)$ measurement. This
close correlation is, at least in part, due to the large uncertainty in
individual overdensity, $\log_{10}(1+\delta_3)$, measures. The bias towards
$\Delta_3 > \log_{10}(1 + \delta_3)$ is a product of the color--density
relation and the inclusion of red--sequence galaxies in the measurement of
galaxy overdensities (see \S \ref{sec_data_environ}).

\section{The Residual Dependence of Environment on Metallicity}
\label{sec_res2}

\begin{figure*}[tb]
\plotone{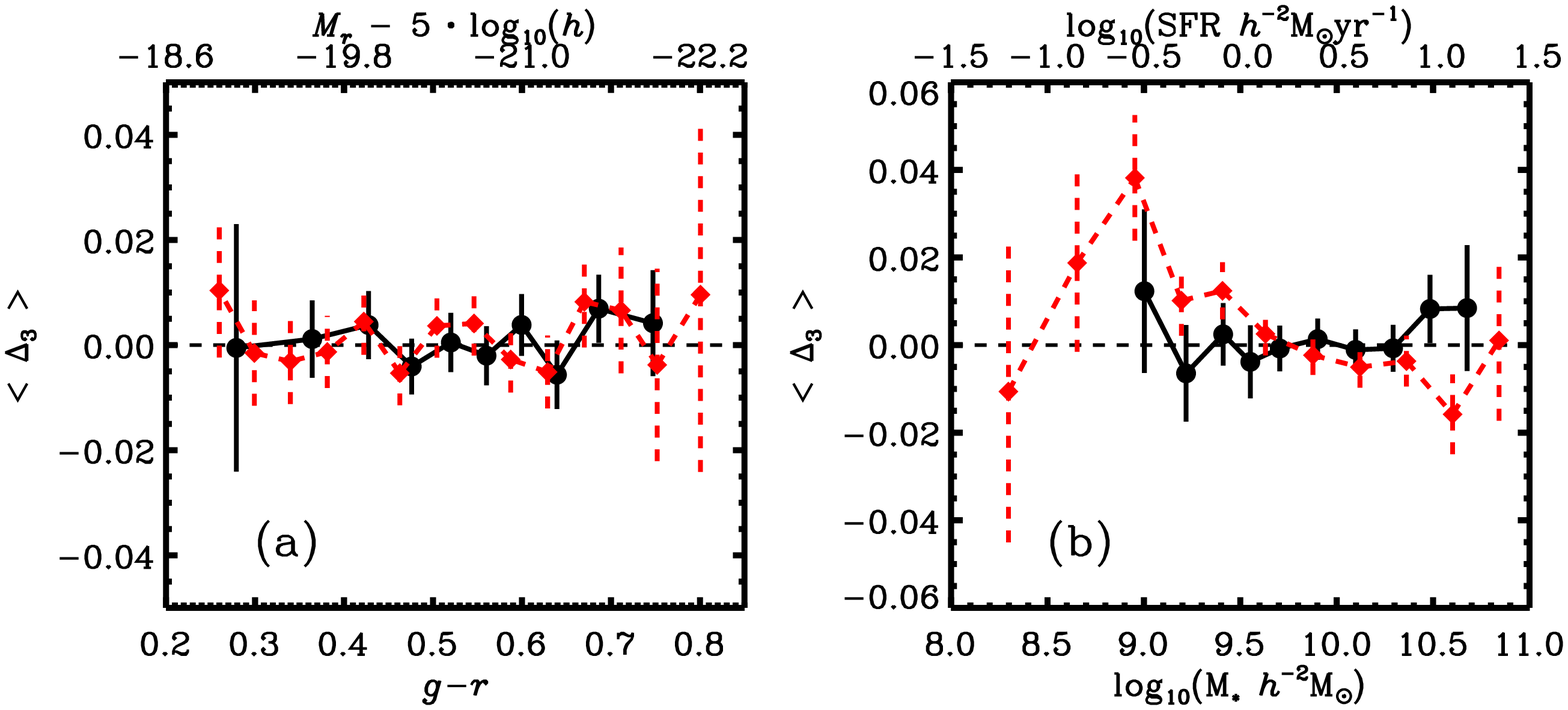}
\caption{(\emph{Left}) The dependence of mean residual environment,
  $\Delta_3$, on rest--frame color and absolute magnitude, as given by the
  black circles plus solid line and red diamonds plus dashed line,
  respectively. (\emph{Right}) Similar to the plot on the left, but for
  stellar mass, ${\rm M}_{*}$, (black circles and solid line) and
  star--formation rate (red diamonds and dashed line). After removing the
  mean dependence of environment on color and luminosity, we find no
  significant residual trend with color, luminosity, stellar mass, or
  star--formation rate.}
\label{resid_fig1}
\end{figure*}

By studying the dependence of residual environment, $\Delta_3$, on various
galaxy properties, we can determine whether there is any excess trend with
environment beyond that contained in the color--luminosity--environment
relation. As a sanity check, in Figure \ref{resid_fig1}a we examine the
dependence of mean $\Delta_3$ on color or absolute magnitude, and confirm
that there is no trend with these properties, as expected. We likewise test
for any dependence of residual environment on stellar mass or
star--formation rate (see Figure \ref{resid_fig1}b). 

We find no signigicant trend of $\Delta_3$ with color, luminosity, stellar
mass, or SFR. This result is clearly to be expected for color and
luminosity, by construction. Given the relatively tight relationship
between the combination of $g-r$ and $M_r$ with ${\rm M}_{*}$
\citep[e.g.,][]{kauffmann03a, cooper07}, it is also not surprising to find
no residual trend with stellar mass, as an additional test. When using the
stellar mass values of \citet{kauffmann03a}, which were derived from fits
to stellar absorption features in the SDSS spectra rather than computed
directly from the SDSS photometry (see \S \ref{sec_data_photo}), we find a
similar lack of any trend. For star--formation rate, which exhibits a
weaker correlation with absolute environment, $\log_{10}(1+\delta_3)$, we
find no evidence for a relationship with residual environment, $\Delta_3$,
much like the lack of secondary environment--dependencies on surface
brightness or S\'{e}rsic index found by \citet{blanton05a}.

\begin{figure}[h!]
\plotone{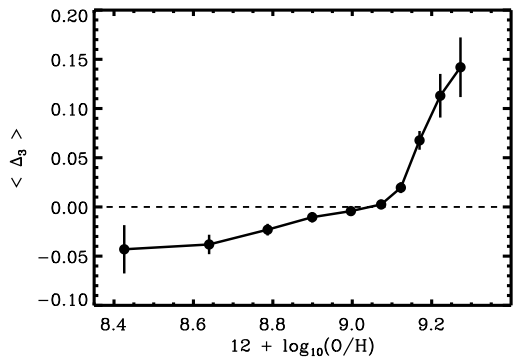}
\caption{The dependence of mean residual environment, $\Delta_3$, on
  metallicity. We find a strong trend with metal--rich galaxies being
  found, on average, in regions of higher overdensity relative to galaxies
  of like color and luminosity.}
\label{resid_fig2}
\end{figure}

Turning our attention towards metallicity, we examine the dependence of
mean residual environment, $\Delta_3$, on gas--phase oxygen abundance; as
shown in Figure \ref{resid_fig2}, there is a striking trend such that more
metal--rich galaxies typically reside in more overdense environments
relative to galaxies of like color and luminosity (i.e., of like stellar
mass). While the residual environment statistic, $\Delta_3$, has no
relationship with color, luminosity, stellar mass, or SFR, it is strongly
related to metallicity. In particular, this trend seems to be most
significant among the most metal--rich galaxies $(12 + \log_{10}({\rm
  O}/{\rm H}) > 9.1)$.

\begin{figure}[h]
\plotone{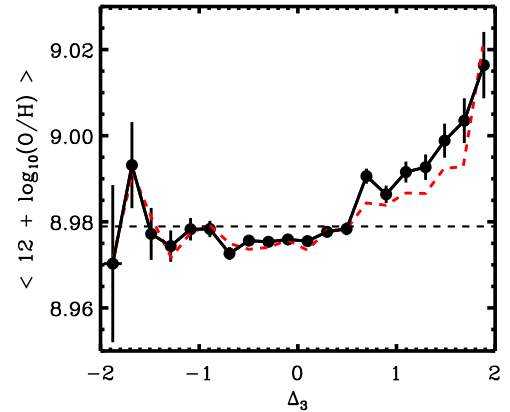}
\caption{The dependence of mean (black points and solid line) and median
  (red dashed line) metallicity on the residual environment, $\Delta_3$. A
  strong correlation is found between residual environment and metallicity
  in overdense regions. This trend is evident, despite smearing effects
  related to the relatively large uncertainty in individual environment
  measures (see text) and the statistical dominance of galaxies with
  metallicities at $12 + \log_{10}({\rm O}/{\rm H}) \sim 9$, where the
  metallicity--environment relation is weak (see Fig.\
  \ref{resid_fig2}). Note that in this figure the median relation has been
  offset by $-0.03$ in $12 + \log_{10}({\rm O}/{\rm H})$ to facilitate
  display.}
\label{resid_fig3}
\end{figure}

Given that the residual environment closely traces the absolute overdensity
measurement (see Fig.\ \ref{diff_fig}), it is interesting to examine this
residual metallicity--environment relation from the opposite
perspective. Figure \ref{resid_fig3} shows the dependence of mean
gas--phase oxygen abundance on the residual environment within the SDSS
star--forming sample. While studying mean relations from this perspective
is physically intuitive, binning galaxies according to environment
(residual or absolute) introduces significant correlation between
neighboring environment bins, due to the significant uncertainties in
measuring local galaxy densities ($\sigma_{\log(1+\delta_3)} \sim 0.5$
versus $\sigma_{12 + \log({\rm O}/{\rm H})} \sim 0.1$), which can therefore
weaken or erase any underlying trends. Despite this smearing effect, we
still find a strong trend, where the mean metallicity increases
dramatically in higher density regions $(\Delta_3 \gtrsim 1)$; this
suggests that the residual metallicity--environment relation is dominated
by phenomena occurring in overdense regions (such as groups and clusters),
rather than underdense environments.

\section{Scatter in the Mass--Metallicity Relation}
\label{sec_scatter}

\begin{figure}[h]
\plotone{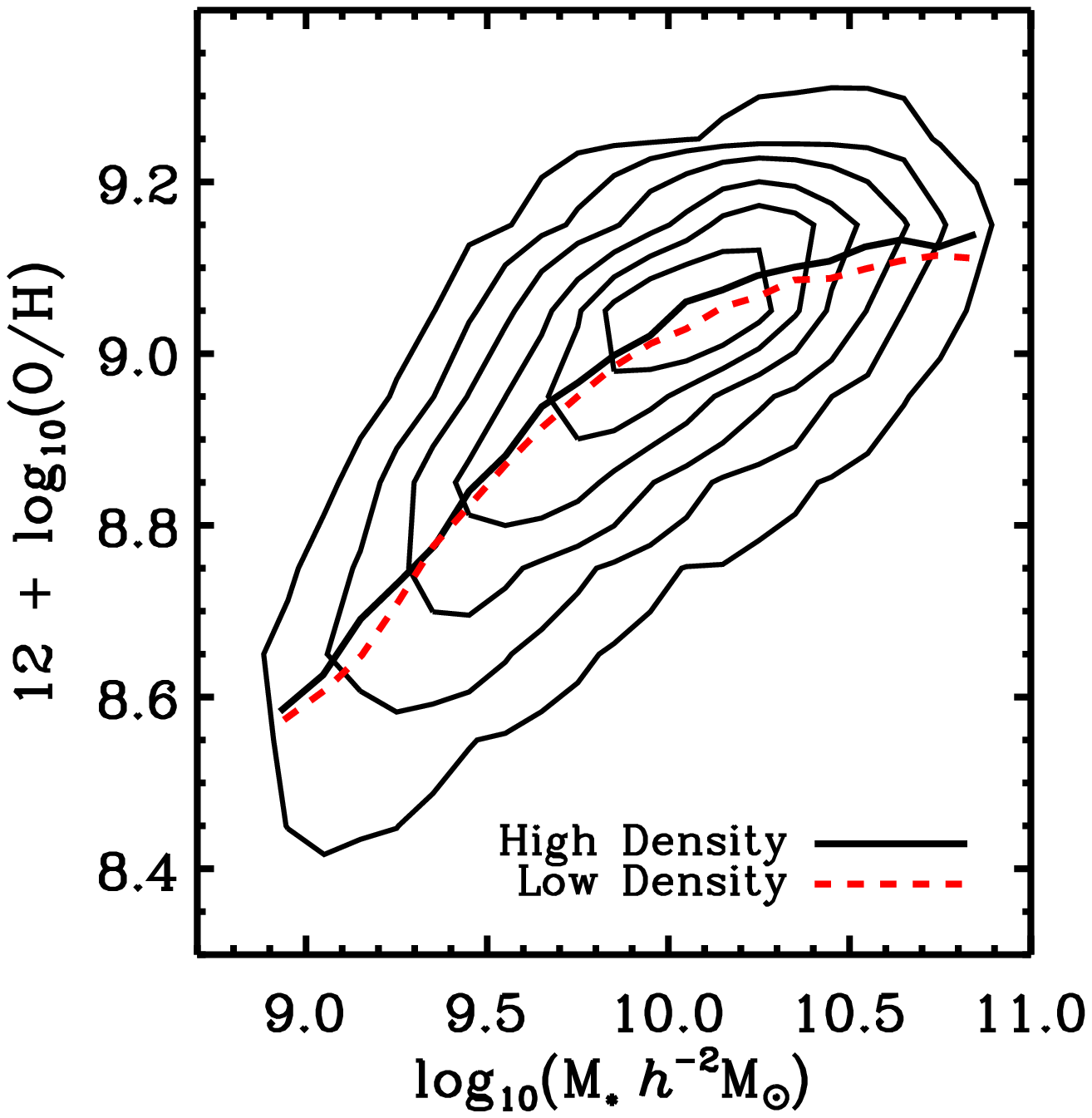}
\caption{We plot the gas--phase oxygen abundance versus stellar mass for
  the star--forming sample and overplot the fits to the mass--metallicity
  relation in the extreme quintiles of the residual overdensity
  distribution. The contours correspond to galaxy numbers of $N_{\rm
    galaxy} = 50, 200, 500, 1000, 1500, 2000$, while the solid black and
  dashed red lines show the median mass--metallicity relation for galaxies
  residing in high-- and low--density regions, respectively. The lines
  follow the median metallicity values, computed in discrete bins of
  stellar mass. At all masses, the median metallicity of galaxies in
  high--density regions is greater than that of galaxies in low--density
  regions.}
\label{mz_fig}
\end{figure}

The excess correlation between metallicity and environment, beyond that
contained in the color--luminosity--environment relation (or stellar
mass--environment relation), strongly suggests that the shape or
normalization of the mass--metallicity relation must depend on local galaxy
environment. This suggestion is confirmed in Figure \ref{mz_fig}, where we
show fits to the mass--metallicity relation, computed using galaxies in the
extreme quintiles of the environment distribution. Over the entire range of
stellar masses probed by the SDSS sample, the mass--metallicity relation is
biased towards higher metallicities in higher--density regions.

While Figure \ref{mz_fig} clearly illustrates the environment dependence of
the mass--metallicity relation, showing an offset towards higher
metallicity in higher--density regions, it does not \emph{quantify} the
level to which environment contributes to the scatter in this fundamental
relationship. To this end, we examine the correlation between environment
and the residual metallicity, $\Delta_{({\rm O}/{\rm H})} = 12 +
\log_{10}({\rm O}/{\rm H}) - f({\rm M}_{*})$, measured relative to the
median mass--metallicity relation, $f({\rm M}_{*})$, as determined by the
full star--forming sample.

\begin{figure}[h]
\plotone{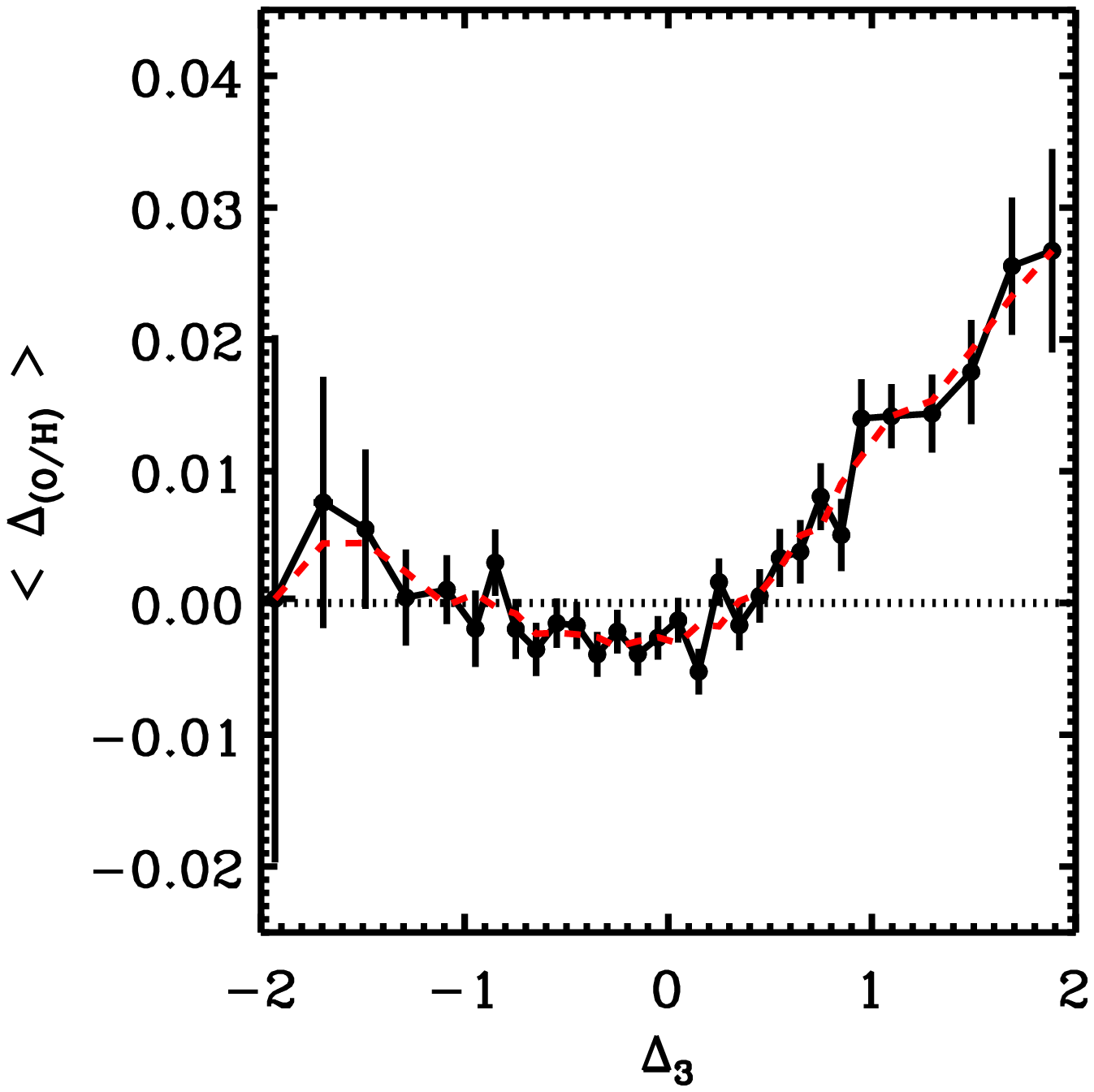}
\caption{The median residual metallicity, relative to the median
  mass--metallicity relation, as a function of environment. We find a
  significant offset in metallicity (relative to the median
  mass--metallicity relation) as a function of galaxy overdensity. The
  dashed red line is the smoothed relation used to compute $\epsilon$. Note
  that the dependence of the mean residual metallicity on environment
  closely follows the relation shown for the median residual metallicity.}
\label{mz_fig2}
\end{figure}

As shown in Figure \ref{mz_fig2}, the average residual metallicity
exhibits a clear dependence on environment, such that galaxies in overdense
regions are biased towards higher metallicities than galaxies of like
stellar mass. This result is effectively a rephrasing of the trend shown in
Fig.\ \ref{resid_fig3} and Fig.\ \ref{mz_fig}, except that in this form we
are able to subtract the average offset in the mass--metallicity relation
due to environment, yielding a quantity
\begin{equation}
  \epsilon = 12 + \log_{10}({\rm O}/{\rm H}) - < \Delta_{({\rm
      O}/{\rm H})}[\Delta_{3}] >,
\end{equation}
which gives the metallicity corrected for the observed environment
dependence. 

Subtracting (in quadrature) the measured scatter in the mass--$\epsilon$
relation from the scatter in the mass--metallicity relation, we find that
environment is correlated with $\gtrsim \! 15\%$ of the observed scatter in
the mass--metallicity relation. This environment--dependence is evident,
with comparable strength, at all stellar masses. As discussed in \S 4 and
\S 5, the relatively large uncertainties in the environment measurements
can smear out the underlying correlation between metallicity and
environment, thereby weakening the measured contribution of environment to
the scatter in the mass--metallicity relation. Thus, local environment is
correlated with \emph{at least} $15\%$ of the observed scatter, which
represents a non--negligible contribution to the total intrinsic scatter.

\begin{figure}[h]
\plotone{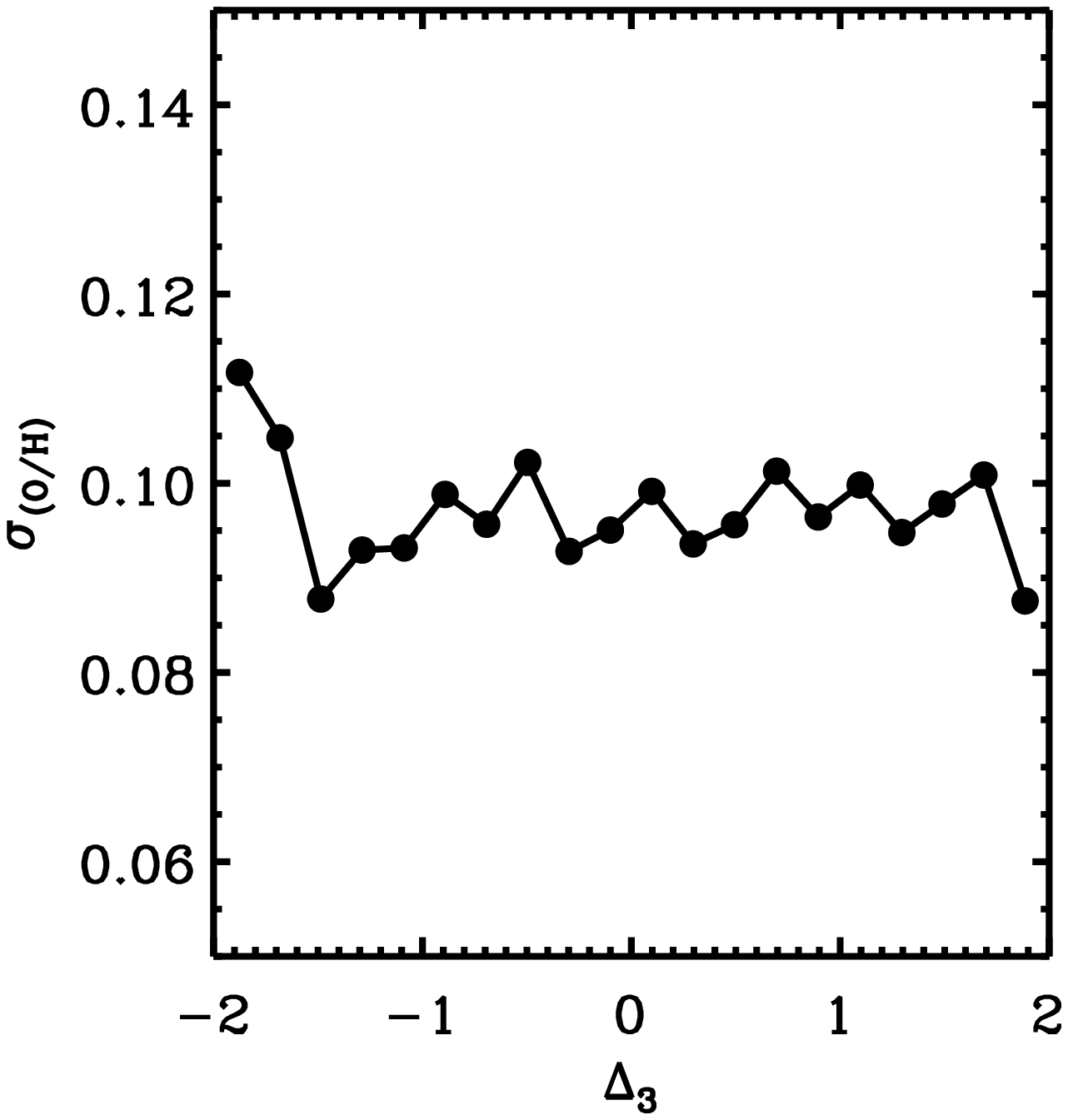}
\caption{The root--mean-square of the devations in the mass--metallicity
  relation as a function of environment. The intrinsic scatter (or
  ``puffiness'') of the mass--metallicity relation shows no variation with
  environment. Note that the errors on $\sigma_{({\rm O}/{\rm H})}$ are
  smaller than the data points, since each bin contains $> \! 500$
  galaxies.}
\label{mz_fig3}
\end{figure}

While we find a significant offset in the normalization of the
mass--metallicity relation in different environments, we do not detect any
environment--dependent variation in the intrinsic scatter. As shown in
Figure \ref{mz_fig3}, the measured root--mean-square (RMS) scatter in the
mass--metallicity relation is independent of environment, at a constant
level of roughly $\sigma_{{\rm O}/{\rm H}} \sim 0.1$. This suggests that
whatever is dominating the intrinsic scatter in the mass--metallicity
relation is independent of local galaxy overdensity.

\section{Discussion}
\label{sec_disc}

\subsection{Potential Selection Effects}

While we utilize the relatively conservative line--diagnostic criteria of
\citet{kauffmann03b} for excluding AGN from our sample, any significant
amount of contamination from AGN emission in the integrated galaxy spectra
could potentially impact the oxygen abundance measurements, biasing them
towards high (or low) metallicity. If AGN are strongly correlated with a
given environment (e.g., if they are preferrentially found in high--density
regions), then the metallicity--density relation could be (at some level) a
product of this underlying AGN--environment correlation. Of particular
interest is the relationship between Low Ionization Nuclear Emission--line
Regions \citep[LINERs,][]{heckman80} and environment as low--level AGN such
as LINERs are more likely to contaminate the star--forming sample than
their more powerful Seyfert counterparts.

Using the SDSS data set, several studies of the relationship between AGN
activity and environment have uncovered no significant correlation between
low--level AGN activity and local galaxy density. For example,
\citet{miller03} found that the fraction of AGN shows no variation with
environment within the SDSS early data release \citep{stoughton02}, a
result supported by later work using the larger DR4 data set
\citep{sorrentino06}. While analysis by \citet{montero08} shows that the
fraction of LINERs and Seyferts on the red sequence is potentially lower in
high--density environments locally, this result may not be representative
of the environments of LINERs in the blue cloud (i.e., among the
star--forming population). In partial agreement with the results of
\citet{montero08}, \citet{kauffmann04} conclude that the fraction of
galaxies hosting a powerful ($L$[O {\small III}]\ $> 10^7 L_{\sun}$) AGN
decreases in high--density environments. However, these are not the AGN
that are likely to contaminate our star--forming sample. For low--level AGN
activity, there is no evidence for a correlation with environment and thus
it is unlikely that contamination from AGN would contribute to the observed
correlation between metallicity and environment in our sample.

As shown by several studies of star--forming galaxies in the local
Universe, there is a correlation between gas--phase oxygen abundance and
galaxy morphology, such that more bulge--dominated systems are typically
more metal--rich \citep[e.g.,][]{vila92, zaritsky93, zaritsky94}. An
analogous trend is found when studying stellar metallicities among a more
diverse galaxy population \citep{gallazzi08}. Any relationship between
metallicity and environment separate from that observed with stellar mass
could therefore be a derivative of the well--known morphology--density
relation \citep[e.g.,][]{davis76, dressler80}. 

Given the strong correlations between luminosity, color, and morphology on
the blue cloud, the existence of a significant correlation between residual
environment, $\Delta_3$, and morphology is unlikely when none is found with
luminosity or rest--frame color. However, we investigate this possibility
using the S\'{e}rsic indices of \citet{blanton03a, blanton05a}. While the
S\'{e}rsic index is a measure of morphology derived from the fit of only a
single component to the galaxy's radial profile (versus bulge--disk
decomposition, for example), we find no dependence of mean residual
environment on S\'{e}rsic among our sample. Furthermore, recent analysis of
star--forming galaxies in the SDSS found that the mass--metallicity
relation shows no dependence on bulge fraction \citep{ellison08a}. Plus, as
stated in \S \ref{sec_intro}, \citet{tremonti04} found no correlation
between the scatter in the mass--metallicity relation and galaxy
concentration. Thus, we conclude that the portion of the scatter in the
mass--metallicity relation correlated with environment is not attributable
to variations in galaxy morphology.

\subsection{Theoretical Interpretation}
\label{sec_disc_1}
As discussed in \S \ref{sec_intro}, gas--phase metallicity and its
relationship with stellar mass within the star--forming population is
directly connected to feedback associated with star formation, as metals
are added to the ISM via supernovae and as gas is ejected via outflows and
accreted from the surrounding intergalactic medium (IGM). The presence of
outflows in star--forming galaxies has been supported by a variety of
observations \citep[e.g.,][]{lehnert96, frye02, weiner08}, but the physics
of this feedback mechanism remains poorly understood.

In an attempt to explain the mass--metallicity relation, early feedback
models \citep[e.g.,][]{dekel86, cole91, dekel03} employed energy--driven
winds, powered by supernovae explosions \citep{larson74}, to expel metals
from low--mass galaxies. Such models, however, fail to include the role of
winds in more massive systems (${\rm M}_{*} \gtrsim 10^{10}\ {\rm
  M}_{\sun}$), while observational work has shown outflows to be common at
galaxy stellar masses of $\gtrsim 10^{11}\ {\rm M}_{\sun}$
\citep[e.g.,][]{shapley03, rupke05, weiner08}.  

In contrast, the models of \citet{springel03a} incorporate winds at all
mass scales, but their simple prescription relies on winds of a constant
velocity \citep[$484\ {\rm km}/{\rm s}$,][]{springel03b}, independent of
galaxy mass. In disagreement with this approach, recent observations by
\citet{martin05} show outflow velocities to scale approximately linearly
with circular velocity (i.e., increase with $M_{*}$). Furthermore, simple
wind approximations such as that of \citet{springel03b} fail to reproduce
some properties of the IGM at higher redshift \citep[e.g., underpredicting
metal enrichment,][]{aguirre05} and the mass--metallicity relation at $z
\sim 2$ \citep{finlator07}.

Recent work by \citet{finlator07} has ventured to take a more detailed
approach to modeling the feedback in star--forming galaxies \citep[see
also][]{oppenheimer06}. In their model, outflows are pushed by
momentum--driven winds \citep{mqt05}, where momentum is deposited into the
ISM by coupling with the radiation from star formation through dust
absorption and where the wind speed scales with the galaxy's circular
velocity. Rather than assuming a wind that is driven in all directions
\citep[such as that of][]{springel03a}, \citet{finlator07} model polar
outflows with constrained opening angles ($\sim \! 45^{\circ}$) such that
the resulting outflows much more closely imitate those observed locally
\citep[e.g.,][]{veilleux05}.

In addition to assuming a wind speed that scales linearly with rotational
speed, the \citet{finlator07} model assumes that the mass--loading factor
--- the rate of mass ejection divided by the star--formation rate --- is
inversely related to the circular velocity. These scaling relations evolve
naturally for momentum--driven winds \citep{mqt05} and are in rough
agreement with results from other detailed feedback models
\citep[e.g.,][]{kobayashi07, brooks07}. Within this theoretical framework,
the gas--phase metallicity at any epoch depends on (\emph{i}) the mean
metallicity of accreted gas and (\emph{ii}) the mass--loading factor
\citep[see Equation 20 of][]{finlator07}.

In this model, the observed trends between metallicity and environment
would require either higher enrichment of the gas flowing into galaxies in
overdense regions and/or lower mass--loading factors in high--density
environments. There are many environment--dependent physical mechanisms
that could yield the former; for instance, supernova feedback from evolved
stars associated with intragroup or intracluster light will directly dump
metals (in particular, oxygen) into the IGM about galaxies in the
highest--density environments. In addition, galaxy mergers, harassment, and
ram--pressure stripping in groups and clusters can strip enriched gas from
member and infalling galaxies, thereby inflating the metal content of the
local gas reservoir relative to the gas supply of roughly primordial
composition that feeds galaxies in the field \citep[e.g.,][]{gunn72,
  moore96, hester06, gnedin98}.

Stripping of gas from cluster members could also contribute to a higher
gas--phase metallicity in extreme environments in a secondary manner. That
is, ram--pressure stripping could remove the outer portion (and therefore
most metal--poor segment) of a galaxy's gas halo. Since the mixing time
(assumed to be the dynamical time) for a disk galaxy is on the order of the
cluster crossing time ($\sim 2$ Gyr), if not stripped this metal--poor gas
would become effectively mixed, thereby reducing the mean metallicity
within the central $\sim 5-10$ kpc (the region sampled by an SDSS fiber).

In the most extreme environments, pressure from the intercluster medium
(ICM) could potentially resist such stripping
\citep[e.g.,][]{babul92}. However, hydrodynamical simulations have found
that the net effect of thermal pressure and ram--pressure stripping on a
cluster member still results in gas being removed from the galaxy,
contributing to the ICM \citep{murakami99}. On the other hand, numerical
and analytical modeling of feedback in isolated galaxies shows that the
ejection of metals from a galaxy's ISM is more likely to occur in regions
of lower pressure \citep[e.g.,][]{silich01, maclow99}. Thus, thermal
pressure (and its impact on the ability to drive an outflow) could account
for the relative decrease in metallicity for galaxies in low--density
environs.

Alternatively, the metallicity--environment relations presented in this
work could also result from variations in the mass--loading factor with
local galaxy density. While the mass--loading factor is, in principal, a
quantity that can be directly observed \citep[e.g.,][]{morganti05},
detailed radio measurements of a galaxy's gas mass are required. Since we
lack the required observations within the SDSS data set, we instead utilize
the SDSS spectroscopic data to look for signatures of variation in outflow
velocity with environment at $z \sim 0.1$. Although, wind speed does not
necessarily provide any information about the amount of mass expelled from
a galaxy, a significant variation in outflow velocity with environment
could be an indication that the net accretion rate (relative to the SFR) is
driving the observed metallicity--environment relations. From co--adding
two sets of spectra including several hundred strongly star--forming (${\rm
  H}\alpha\ {\rm equivalent\ width} > 30$\AA), massive (${\rm M}_{*} >
10^{10}\ {\rm M}_{\sun}$) galaxies, we find no significant variation in the
Na D absorption profile between extreme (low--density and high--density)
environments. Admittedly, our analysis is limited to the most highly
star--forming galaxies, given the low resolution $({\rm R} \sim 1800)$ of
the SDSS spectra.

Another point to consider when searching for physical sources of the strong
relationship between environment and metallicity is that galaxies
populating high--density regions today likely formed early in the first
overdensities. Predictions of early galaxy enrichment
\citep[e.g.,][]{schaye03, dave06} indicate that these overdensities of gas
at high--$z$ would be the most enriched environments, naturally producing a
metallicity--environment relation \citep[see also][]{oppenheimer06}.
Within the model of \citet{finlator07}, however, the gas--phase metallicity
in a galaxy at $z \sim 0.1$ is a product of the recent ($< \! 1$ Gyr)
accretion and star--formation activity, rather than a result of the
integrated star--formation history of the galaxy \citep[see
also][]{dalcanton07}. So while galaxies in high--density environs in the
local Universe generally formed early in cosmic time and in the early
density peaks, metallicity--environment relations imprinted at $z \gtrsim
2$ would not necessarily persist to the present.

\subsection{Comparison to Related Work}
\label{sec_disc_2}

As this paper was being completed, a parallel analysis of the relationship
between metallicity and environment in the SDSS was presented by
\citet{mouhcine07}. Using a very similar data set, drawn from SDSS DR4 and
employing the metallicity measurements of \citet{tremonti04}, they find
that the mass--metallicity relation depends weakly on local
environment. When dividing our sample into discrete bins according to
overdensity, we also find a relatively weak connection between metallicity
and environment at fixed stellar mass (see Fig.\ \ref{mz_fig}); that is,
when plotting the median mass--metallicity relation in discrete bins of
overdensity, we find what appear to be only small variations among
environment regimes, in close agreement with Figure 5 of
\citet{mouhcine07}.

However, studying the relationships between galaxy properties and
environment in this manner is far less sensitive than the techniques
presented herein. Measurements of local galaxy density are inherently
noisier than measures of other galaxy properties, including rest--frame
color, luminosity, stellar mass, and metallicity. Thus, when dividing a
sample by environment, any trends in the data set are smeared out by the
significant correlation between neighboring bins. While \citet{mouhcine07}
conclude that gas--phase oxygen abundance is only weakly dependent on
environment, we have presented evidence to the contrary, showing that the
metallicity--environment relation is roughly equal in strength to the
color--density relation. Furthermore, we find metallicity has a
relationship with environment that is separate from the color--density or
stellar mass--density relations.

In contrast to our work and that of \citet{mouhcine07}, which trace galaxy
environments on $\sim 1$--$2 h^{-1}$ Mpc scales over the full SDSS galaxy
population, the analyses of \citet{kewley06} and \citet{ellison08b} probe a
far more limited range of environments, focusing on the metallicity of
galaxy pairs in the local Universe. Focusing on smaller scales, they find
that galaxy pairs at close (projected) separations ($\lesssim \! 30\
h^{-1}$ kpc) are biased towards lower metallicities. This result is
attributed to inflows of metal--poor gas during the merger or interaction
process, an effect that is also found in simulations \citep{perez06}. The
number of close (projected separations $< \! 100\ h^{-1}$ kpc) pairs,
however, is $\lesssim \! 1\%$ in the SDSS sample \citep[see
also][]{deng06}, and thus such systems cannot be a significant contribution
to the scatter in the mass--metallicity relation. While metallicity may be
lower in close pairs, the dominant metallicity--environment relation moves
towards higher metal enrichment in high--density environments.

\section{Summary and Conclusions}
\label{sec_sum}

Using the measurements of gas--phase oxygen abundance from
\citet{tremonti04} and local galaxy environment from \citet{cooper08}, we
study the relationship between metallicity and environment in a sample of
star--forming galaxies drawn from the SDSS data set. Our principal results
are as follows.

\begin{enumerate}

\item We find a strong metallicity--density relation (see Fig.\
  \ref{metal_fig}b) in the local Universe such that more metal--rich
  galaxies favor regions of higher galaxy overdensity. This relationship
  between metallicity and environment follows (with comparable or greater
  strength) that seen between environment and other fundamental properties
  such as color, luminosity, SFR, or stellar mass.

\item After removing the mean color--luminosity--environment relation from
  the SDSS data set, we find a significant residual relationship between
  environment and metallicity (see Fig.\ \ref{resid_fig2}), suggesting that
  metallicity has a relationship with environment separate from that
  observed with color and luminosity (or with stellar mass). 

\item The residual metallicity--environment trend is largely driven by
  galaxies in high--density regions such as groups and clusters, where the
  local environment may be responsible for impacting the feedback and/or
  gas accretion relative to galaxies of like stellar mass in lower--density
  regions.

\item A non--negligible portion (at least $15\%$) of the scatter in the
  mass--metallicity relation is correlated with local environment.

\end{enumerate}

%%%%%%%%%%%%%%%%%%%%%%%
%%% Acknowledgments %%%
%%%%%%%%%%%%%%%%%%%%%%%

\acknowledgments Support for this work was provided by NASA through the
Spitzer Space Telescope Fellowship Program. C.A.T.\ acknowledges support by
NASA through Hubble Fellowship grant HST--HF--01192.01, awarded by the
Space Telescope Science Institute, which is operated by AURA Inc.\ under
NASA contract NAS 5--26555. M.C.C.\ would like to thank Michael Blanton,
David Hogg, and Renbin Yan for their assistance in utilizing the NYU--VAGC
data products. This work benefited greatly from conversations with Kristian
Finlator, Romeel Dav\'{e}, Ben Weiner, and Dennis Zaritsky.

Funding for the SDSS has been provided by the Alfred P.\ Sloan Foundation,
the Participating Institutions, the National Science Foundation, the
U.S.\ Department of Energy, the National Aeronautics and Space
Administration, the Japanese Monbukagakusho, the Max Planck Society, and
the Higher Education Funding Council for England. The SDSS Web Site is
http://www.sdss.org/.

The SDSS is managed by the Astrophysical Research Consortium for the
Participating Institutions. The Participating Institutions are the American
Museum of Natural History, Astrophysical Institute Potsdam, University of
Basel, University of Cambridge, Case Western Reserve University, University
of Chicago, Drexel University, Fermilab, the Institute for Advanced Study,
the Japan Participation Group, Johns Hopkins University, the Joint
Institute for Nuclear Astrophysics, the Kavli Institute for Particle
Astrophysics and Cosmology, the Korean Scientist Group, the Chinese Academy
of Sciences (LAMOST), Los Alamos National Laboratory, the
Max--Planck--Institute for Astronomy (MPIA), the Max--Planck--Institute for
Astrophysics (MPA), New Mexico State University, Ohio State University,
University of Pittsburgh, University of Portsmouth, Princeton University,
the United States Naval Observatory, and the University of Washington.

%%%%%%%%%%%%%%%%%%%%
%%% Bibliography %%%
%%%%%%%%%%%%%%%%%%%%
%\bibliographystyle{apj}
%\bibliography{apj-jour,environ_metal_refs}

\end{document}